\documentclass[onecolumn,superscriptaddress,showpacs,pra]{revtex4}
\usepackage{epsfig}
\usepackage{epstopdf}
\usepackage{delarray}
\usepackage{amsmath, amssymb}
\usepackage{bm}
\usepackage{graphicx}% Include figure files
\usepackage{placeins}
\usepackage{color}
\begin{document}
\title{The Solution of the Long-Wave Equation for Various Nonlinear Depth and Breadth Profiles in the Power-Law Form}

\author{Cihan Bay\i nd\i r}
\email{cbayindir@itu.edu.tr}
\affiliation{Associate Professor, Engineering Faculty, \.{I}stanbul Technical University, 34469 Maslak, \.{I}stanbul, Turkey. \\
						 Adjunct Professor, Engineering Faculty, Bo\u{g}azi\c{c}i University, 34342 Bebek, \.{I}stanbul, Turkey. \\
						 International Collaboration Board Member, CERN, CH-1211 Geneva 23, Switzerland.}

\author{Sofi Farazande}
\affiliation{Graduate Student, Engineering Faculty, Bo\u{g}azi\c{c}i University, 34342 Bebek, \.{I}stanbul, Turkey.}

%\date{\today}

\begin{abstract}
%% Text of abstract
Long waves bring many important challenges in the ocean and coastal engineering, including but are not limited to harbor resonance and run-up. Therefore, understanding and modeling their dynamics is crucially important. Although their dynamics over various types of geometries are well-studied in the literature, the study of the geometries with power-law variations remains an open problem in this setting. With this motivation, in this paper, we derive the exact analytical solutions of the long-wave equation over nonlinear depth and breadth profiles having power-law forms given by $h(x)=c_1 x^a$ and $b(x)=c_2 x^c$, where the parameters $c_1, c_2, a, c$ are some constants. We show that for these types of power-law forms of depth and breadth profiles, the long-wave equation admits solutions in terms of Bessel functions and Cauchy-Euler series. We also derive the seiching periods and resonance conditions for these forms of depth and breadth variations. Our results can be used to investigate the long-wave dynamics and their envelope characteristics over equilibrium beach profiles, the effects of nonlinear harbor entrances and angled nonlinear seawall breadth variations in the power-law forms on these dynamics, and the effects of reconstruction, geomorphological changes, sedimentation, and dredging to harbor resonance, to the shift in resonance periods and to the seiching characteristics in lakes and barrages. 

\end{abstract}
\maketitle

\section{Introduction}
In ocean and coastal modeling and engineering fields, one of the most challenging problems for the designs is to suppress the effects of long-waves which lead to very large period oscillations that are very hard to dissipate, large amounts of wave runup, overwash, and inundation. The literature on this well-studied subject is vast \cite{Lamb, Chrystal, Wilson, Goldsbrough, DeanDal, Mei}. Linear long-wave equation is first published by Lamb in the first edition of his masterpiece Hydrodynamics book in 1879 \cite{Lamb}, and in an approximate form by Chrystal in 1905 \cite{Chrystal}. Since then these equations are analyzed extensively \cite{Lamb, Chrystal, Wilson, Goldsbrough, DeanDal, Mei}. While the majority of these studies are focused on the analysis of long-wave equations for different estuary, bay, and harbor geometries, there are also many studies on their interaction with sediments, beaches, barred beaches, and islands such as those discussed in \cite{Munk, Lettau, Bowen, HigginsStewart, Suhayda, SmithSprinks, Kirby1981, Eckart, Ball67}.

Many investigations of the standing long-waves with oblique incidence to the beach have been conducted for long by the scientific community. Multiple bar formation due to the response of loose sands under such standing waves is one of the observed and studied phenomena. The solution for the long-wave equation for the reflection of normally incident, non-breaking waves coming from a horizontal beach slope has been given by Lamb \cite{Lamb}, and the motion of suspended sediments towards the antinodes of such waves is demonstrated by Lettau \cite{Lettau} and Noda \cite{Noda}. Bowen \cite{Bowen} has also predicted the accumulation of sediments under standing wave nodes and antinodes which is shown to impel the linear offshore bar formation.

Munk \cite{Munk} has named the ``surf beat" the occurrence of long-period progressive waves on the beach, a phenomenon which has been observed for the first time back then. Longuet-Higgins and Stewart \cite{HigginsStewart} have analyzed the mechanism of wave groups of incident wind waves driving a second-order wave of long-period that is traveling at the wind wave group velocity. Gallagher \cite{Gallagher} has studied the surf beat generation by non-linear interaction of 37 wind waves. Long-wave resonance and their scattering around circular islands have also been studied by Smith and Sprinks \cite{SmithSprinks}, and by Summerfield \cite{Summerfield} with a numerical approach. Eckart \cite{Eckart} studied the surface waves where the water depth is variable, and Ball \cite{Ball67} made an analysis on the behavior of waves of different orders moving in various directions.

One of the interesting and crucial phenomena studied within the frame of the long-wave equations are the edge waves \cite{Kirby1981, Ursell}. Due to the vanishing depth related singularity at the shoreline, edge wave solutions are studied using the Frobenius series and solutions for some cases turned out to be in the form of the Laguerre polynomials \cite{Kirby1981}. It is also demonstrated that coastal edge waves can be excited in the infragravity spectral range with a limited analysis including only low edge-wave mode numbers \cite{Gallagher}. These results were extended to edge waves with high mode numbers, by Bowen and Guza \cite{BowenGuza}. Infragravity waves with offshore surface profiles similar to the Bessel function solution were shown to be existent by the fieldwork conducted by Suhayda \cite{Suhayda} in which it is also shown that energy spectrum and phase difference in standing waves are shown to be coherent with the linear long-wave theory developed by Lamb \cite{Lamb}. Some nonlinear theories have also been applied to edge waves to investigate the effects of wave interactions \cite{Ursell, Stokes, Huntley, HolmanBowen, GuzaInman}.

Long-waves are also commonly used in the field of sound and optical engineering \cite{Auld, Lakinacousticresonator, LigeometryResonator, Chevalieropticalresonator}. The sound and optical resonators are tuned to resonate at specific frequencies of the excitations. To model such phenomena with very different and complex geometries, long-wave equations and some of their extensions are commonly employed.

Long-wave equations have well-known solutions for various simple geometries such as uniform depth \cite{Lamb, Wilson, Goldsbrough, DeanDal, Mei}, linearly varying depth or linearly varying breadth (pie-shaped beach) \cite{ Lamb, Wilson, Goldsbrough, DeanDal, Mei}, some combinations of trenches with these types of simple geometries \cite{Wilson}, parabolic and elliptical depth variations \cite{Lamb, Wilson, Goldsbrough}. However, to our best knowledge, long-wave equations are not solved for the geometries where the depth and/or breadth profiles have nonlinear power-law variations.
A special case of such variations in the ocean environment is the equilibrium beach profiles. With this motivation, we derive the exact analytical solutions of the long-wave equation for such profiles in this paper. The solutions are obtained in terms of Bessel-Z Functions and Cauchy-Euler Series, the harbor resonance and seiching hydrodynamics are analyzed for such depth and breadth variations. Long-wave solutions, harbor resonance conditions, and seiching periods derived for such profiles are compared with their well-known counterparts existing in the literature. The uses, applicability, and limitations of the proposed approach, our findings, and possible future research directions are discussed.

\section{\label{sec:level2}Solution of the Long-Wave Equation}

\subsection{\label{sec:level1} Review of the Long-Wave Equation}
Linear long-wave equation is first derived by Lamb \cite{Lamb} and in a simplified form by Chrystal \cite{Chrystal}, and extensively studied by many researchers since then \cite{Lamb, Chrystal, Wilson, Goldsbrough, DeanDal, Mei}. The long-wave equation is given by 
\begin{equation}
\frac{g}{b(x)} \frac{d}{dx}\left[b(x)h(x) \frac{d\eta(x)}{dx} \right]+\sigma^2 \eta=0,
\label{eq01}
\end{equation}
where $g$ denotes the gravitational acceleration, $\sigma=2 \pi /T$ denotes the angular frequency and $T$ denotes the period of the long-wave, $h(x)$ denotes the depth (bathymetry) function and $b(x)$ denotes the breadth (width) function of the estuary. After the solution for $\eta(x)$ are obtained, transient solutions can be constructed by using the time harmonics, i.e. $\widehat{\eta}(x,t)=\eta(x) \cos(\sigma t)$. It is well-known that if linearly varying profiles are used for either the depth function, $h(x)$, or the breadth function, $b(x)$, Eq.~\ref{eq01} admits solution in terms of the Bessel functions of order zero \cite{Lamb}.

To our best knowledge, the nonlinear variations of the estuary depth and breadth in the power-law forms are not studied in the existing literature. With this motivation, in this paper, we generalize the solutions of the long-wave equation to account for these types of variations of $h(x)$ and/or $b(x)$.

\subsection{\label{sec:level3}Solutions in terms of Bessel Functions}
In order to model the long-waves in an estuary having a power-law form of nonlinear variation of depth and breadth, such variations can be formulated as $h(x)=c_1 x^a$ and $b(x)=c_2 x^c$, where $c_1, c_2, a, c$ are some constants. For these profiles, the long-wave equation given in Eq.~\ref{eq01} becomes
\begin{equation}
 \frac{d}{dx}\left[x^{a+c} \frac{d\eta(x)}{dx} \right]+ \frac{\sigma^2}{c_1g} x^{c} \eta=0.
\label{eq02}
\end{equation}
This equation can be reduced to the Bessel equation using power-law change of variables and transformations \cite{Greenberg}. The solution of this equation can be written as
\begin{equation}
 \eta(x)=x^{\nu / \alpha} Z_{|\nu|}\left(\alpha \sqrt{\left|\frac{\sigma^2}{c_1g} \right|} x^{1/\alpha} \right),
\label{eq03}
\end{equation}
where $\alpha=\frac{2}{2-a}$ and $\nu=\frac{1-a-c}{2-a}$. Here, $Z_{|\nu|}$ denotes Bessel function of first kind and order $|\nu|$, $J_{|\nu|}$, and Bessel function of second kind and order $|\nu|$, $Y_{|\nu|}$, if $\frac{\sigma^2}{c_1g}>0$, and the modified Bessel function of the first kind order $|\nu|$, $I_{|\nu|}$, and the modified Bessel function of the second kind and order $|\nu|$, $K_{|\nu|}$, if $\frac{\sigma^2}{c_1g}<0$.

\begin{figure}[htb!]
\begin{center}
   \includegraphics[width=6.9in]{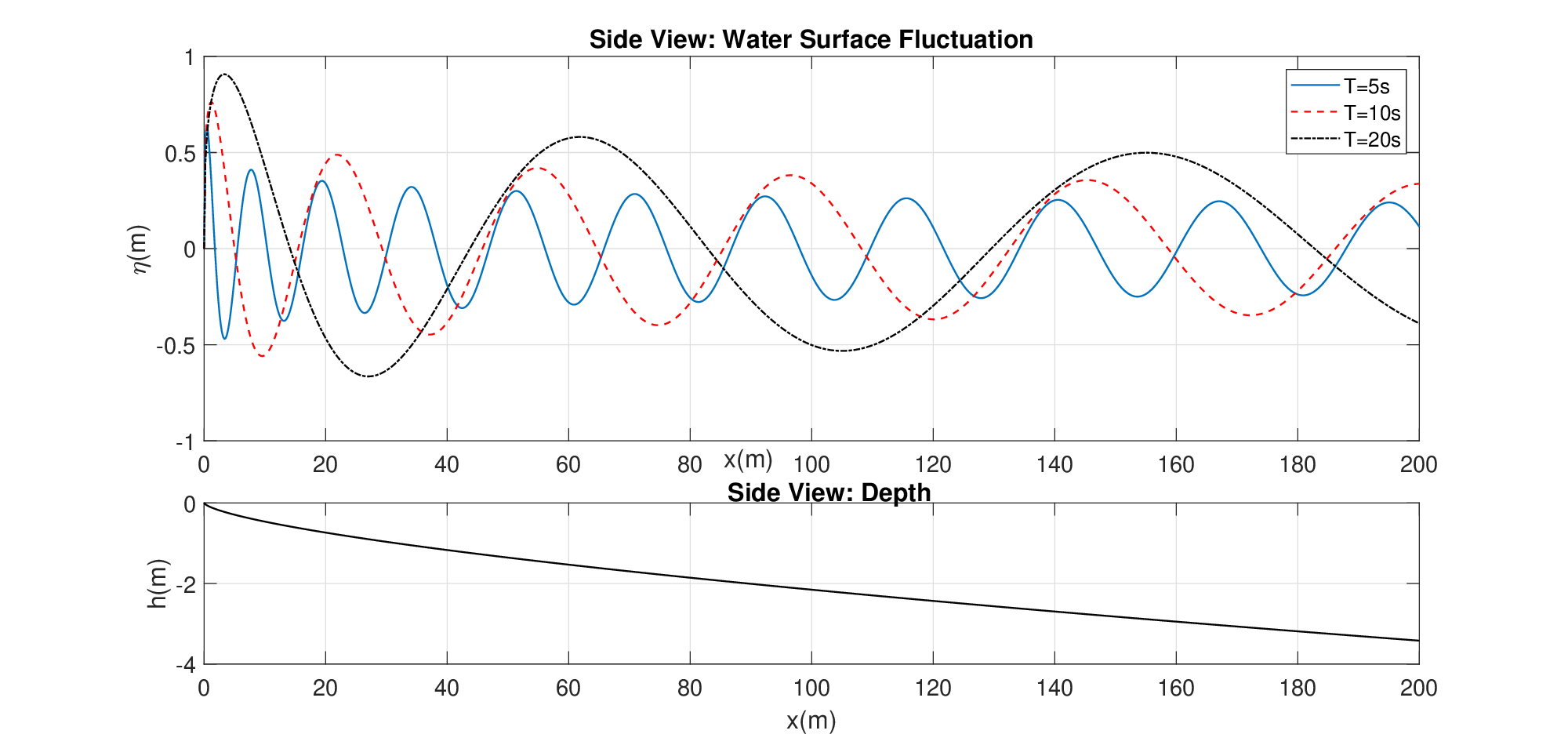}
  \end{center}
\caption{\small Long-wave solution over the equilibrium beach profile with a shape factor of $A=0.1m^{1/3}$ for three different wave periods of $T=5s, 10s, 20s$ at time $t=0$. a) Water surface fluctuation b) Depth variation.}
  \label{fig1}
\end{figure}

{\itshape{Example 1: Equilibrium Beach Profile, $h_{eq}(x)=c_1 x^{a}=A x^{2/3}$, and Uniform Breadth, $b_{uni}(x)=1$:}}
In order to illustrate the possible usage of our findings, we present the solution of the long-wave equation over an equilibrium beach profile as our first example. Using an energy balance approach, Dean \cite{DeanEq} derived the geometric equations that describe the bathymetry of the equilibrium beach profiles which can be summarized as $h_{eq}(x)=A x^{2/3}$. Here, $A$ is the profile scale factor that depends on the sediment size at the sea bottom \cite{DeanEq, DallyDeanDal}. Letting $h_{eq}(x)=c_1 x^{a}=A x^{2/3}$ and $b(x)=1$, we obtain the long-wave profile over an equilibrium beach profile
\begin{equation}
 \eta_{eq}(x)=x^{\nu_{eq} / \alpha_{eq}} Z_{|\nu_{eq}|}\left(\alpha_{eq} \sqrt{\left|\frac{\sigma^2}{A g} \right|} x^{1/\alpha_{eq}} \right),
\label{eq04}
\end{equation}
where the parameters are $\alpha_{eq}=3/2$ and $\nu_{eq}=1/4$. We depict the Bessel function solutions in Fig.~\ref{fig1} for three different periods of $T=5s, 10s, 20s$. Since the condition $\frac{\sigma^2}{c_1g}>0$ is satisfied the Bessel function solutions, not the modified Bessel function solutions, are used. For the solutions plotted in Fig.~\ref{fig1}, a profile scale factor of $A=0.1m^{1/3}$ is used.

As one can see Fig.~\ref{fig1}, the typical Bessel solution behavior, that is the oscillatory behavior and decaying amplitudes is apparent in the offshore direction as expected. However, more strikingly, if the behavior in the nearshore region is investigated, it is clear that the peak water surface elevation occurs at shifted position compared to the shoreline position at $x=0$. The location of peaks can be determined by checking the zeroes of the Bessel function, as explained in more detail in Sec.~\ref{sec:level3}. This shift becomes more prominent for larger wave periods. This shift and the changes in the peak surface elevation can lead to many interesting engineering challenges. First of all, it is clear that the resonance conditions are significantly affected by equilibrium beach profiles and similar topographic features. The harbor lengths, resonant frequencies, and positioning of quay walls and floating units inside harbors and marinas should be carefully examined to take the effect of such depth variations. Additionally, the maximum runup and breaking locations can be affected by such depth and breadth profiles. This would lead to many important lessons in tsunami modeling and swash zone hydrodynamics.

\begin{figure}[htb!]
\begin{center}
   \includegraphics[width=6.9in]{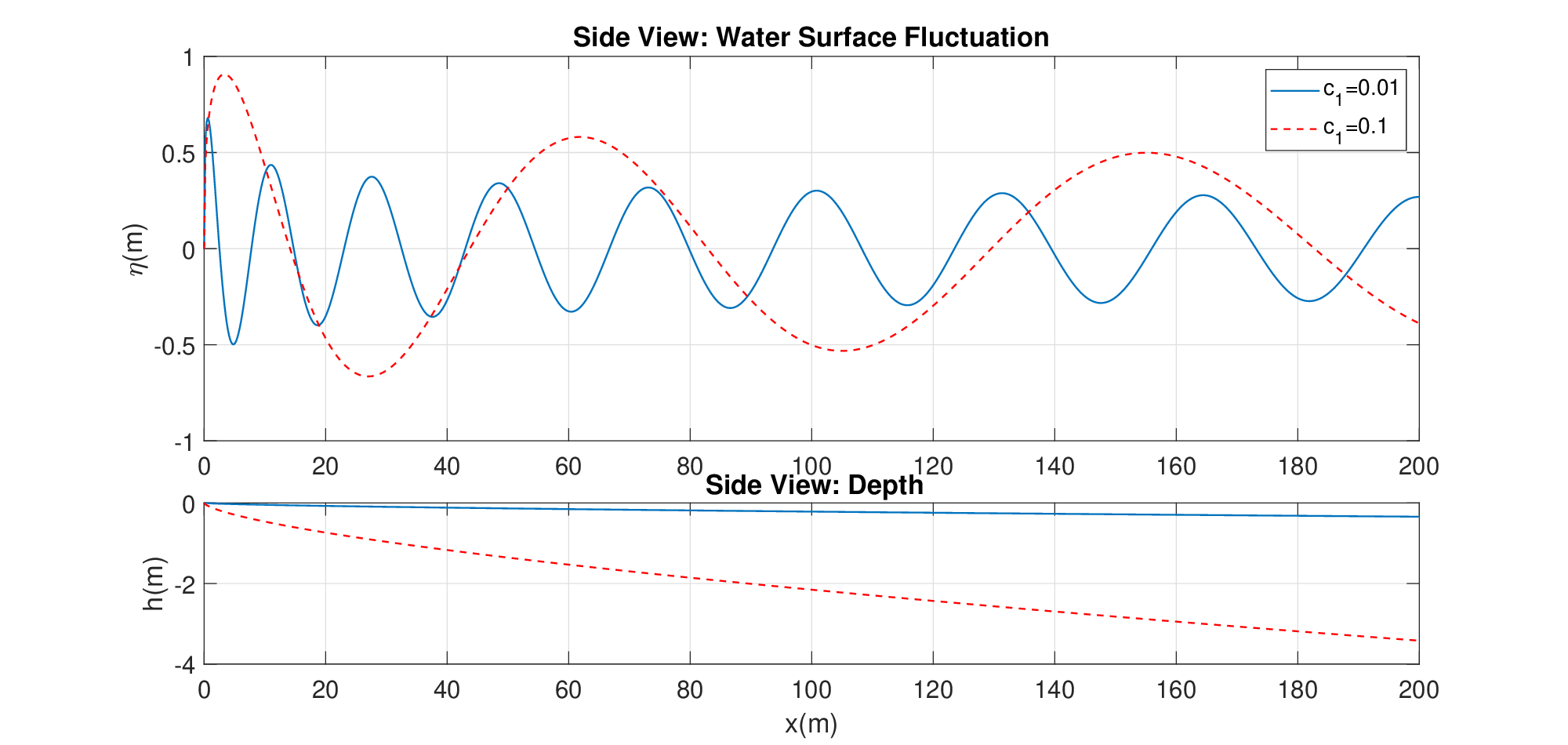}
  \end{center}
\caption{\small Long-wave solution over two different equilibrium beach profiles with shape factors of $c_1=A=0.01m^{1/3}$ and $c_1=A=0.1m^{1/3}$ for a wave period of $T=20s$ at time $t=0$. a) Water surface fluctuation b) Depth variation.}
  \label{Newfig_afterrevision2}
\end{figure}

In order to illustrate the effects of gentle vs. steeper slopes on the solution of the long-wave equation, we depict Fig.~\ref{Newfig_afterrevision2}. In Fig.~\ref{Newfig_afterrevision2}, two different equilibrium beach profiles with $c_1=A=0.01m^{1/3}$ and $c_1=A=0.1m^{1/3}$ are considered and the solution of the long-wave equation corresponding to these parameters are depicted. As one can realize from this figure, on a steeper slope the wavelength and peak amplitudes are larger, as expected. Again such a feature can be beneficial for many engineering purposes such as avoiding harbor resonance or resonating for energy harvesting efficiency. The bathymetric features can be designed and engineered to give desired results.

\begin{figure}[htb!]
\begin{center}
   \includegraphics[width=6.9in]{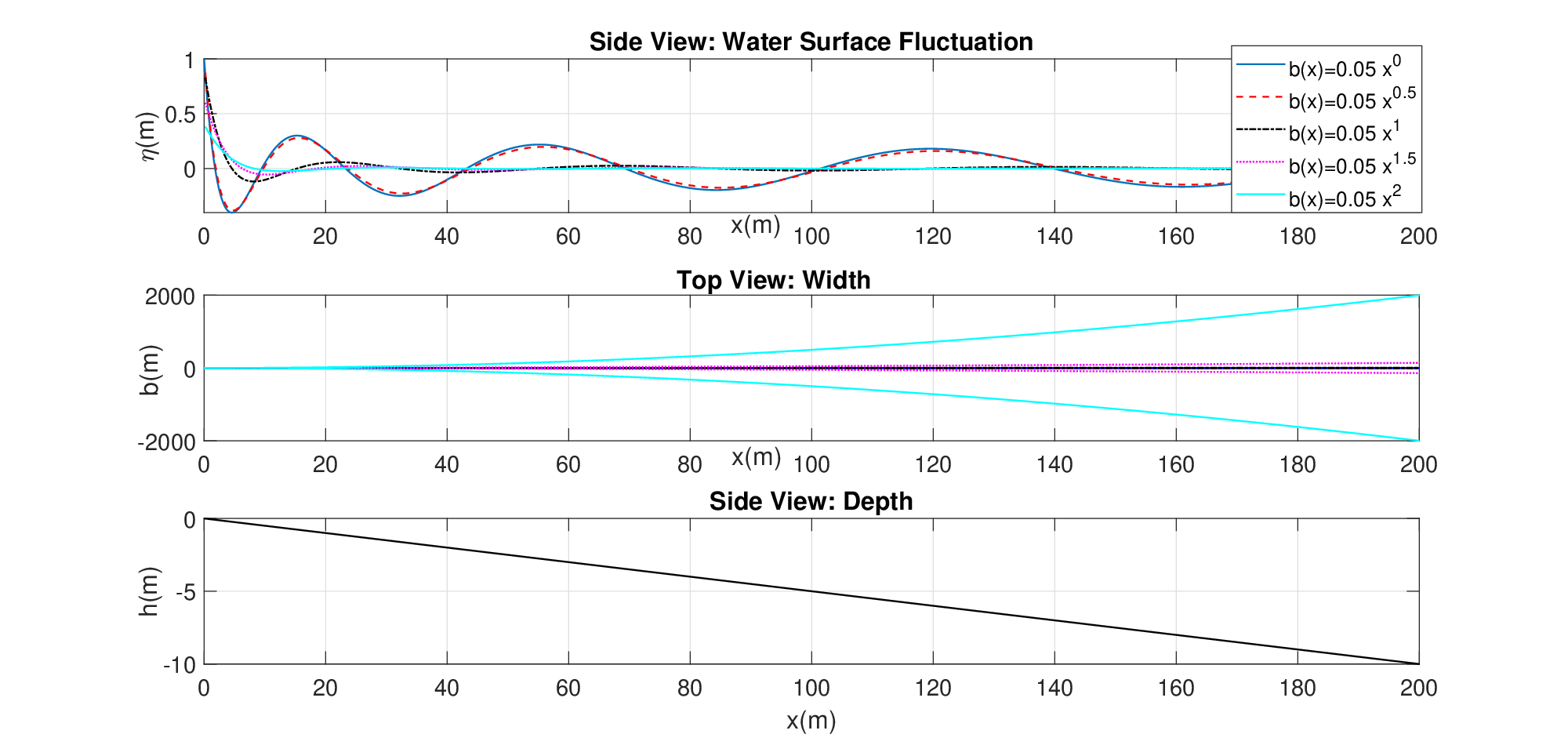}
  \end{center}
\caption{\small Long-wave solution for linearly varying depth and 6 different breadth profiles for a wave period of $T=10s$ at time $t=0$. a) Water surface fluctuation b) Breadth variation c) Depth variation.}
  \label{fig2}
\end{figure}

{\itshape{Example 2: Linearly Varying Bathymetry, $h_{lin}(x)=c_1 x^{a}=0.05 x$, and Various Breadth Profiles-From Uniform to Parabolic, $b_{uni}(x)=c_2 x^{c}=0.05 x^0=0.05$ to $b_{par}(x)=c_2 x^{c}=0.05 x^2$ for various values of $c$:}}
In our second example, we consider the case of linearly varying bathymetry which can be formulated as $h_{lin}(x)=c_1 x^{a}=0.05 x$ if the parameters are selected as $c_1=0.05, a=1$. Together with this linear depth variation, we consider 6 different cases of breadth variation, from uniform to parabolic profiles. The parameters used for the breadth profiles are given in Fig.~\ref{fig2}.

We depict the water surface fluctuation for the various depth-breadth combinations considered in Fig.~\ref{fig2}. As it can be seen in the figure, as the breadth of the estuary becomes larger, the oscillation amplitudes decrease and the wavelength increases. However, in the nearshore region, no shift in the location of the peak of the water surface fluctuation is observed.

\subsection{\label{sec:level3}Solutions in term of Cauchy-Euler Series}

It is well-known that the solutions of the long-wave equation given by Eq.~\ref{eq01} are not described by Bessel or modified Bessel functions for a uniform estuary breadth and a parabolic bathymetry. The solutions for this case are in terms of the Cauchy-Euler series as described in many texts such as \cite{Greenberg, Arfken, Whittaker}. This is also true for the combinations of parabolic bathymetry in the form $h_{par}(x)=c_1 x^{2}$ obtained for $a=2$ with the arbitrary breadth profiles in the power-law nonlinear form of $b(x)=c_2 x^c$. For this case governing equation reduces to
\begin{equation}
 x^2 \eta_{xx}+ (c+2)x\eta_{x}+ \sigma^2/(c_1g)\eta=0.
\label{eq05}
\end{equation}
This equation is known as the Cauchy-Euler equation \cite{Greenberg}. It is possible to solve this equation by seeking a solution in the form of $\eta=Mx^r$ where $M$ is an arbitrary constant \cite{Greenberg}. Seeking a solution in this form and solving characteristic equation gives two roots
\begin{equation}
r_{1,2}= \frac{-c-1 \pm \sqrt{(c+1)^2-4\sigma^2/(c_1g)}}{2}.
\label{eq06}
\end{equation}
The sign of the term inside the square-root in Eq.~\ref{eq06} determines the behavior of the solutions \cite{Greenberg}. It is possible to identify three families of solutions: 
\newline
i) If $(c+1)^2>4\sigma^2/(c_1g)$, then the roots of the characteristic equation $r_{1,2}$ become real, and solution becomes
$\eta=Mx^{r_1}+Nx^{r_2}$ where $M$ and $N$ are arbitrary constants \cite{Greenberg}.

\noindent ii) If $(c+1)^2=4\sigma^2/(c_1g)$, the roots of the characteristic equation $r_{1,2}= (-c-1)/2$ are repeated, and the solution becomes $\eta=Mx^{r_1}+Nx^{r_2} \ln(x)$ \cite{Greenberg}.

\noindent iii) If $(c+1)^2<4\sigma^2/(c_1g)$, the roots of the characteristic equation $r_{1,2}$ become complex, and the solution becomes $\eta=Mx^{r_1}+Nx^{r_2}$. If we re-express the complex roots by $r_{1,2}=\lambda \pm \mu i$, then the solution can be expressed as $\eta=Mx^{\lambda} \cos(\mu \ln(x))+iNx^{\lambda} \sin(\mu \ln(x))$ \cite{Greenberg}.
Due to the linearity of the governing equation, the superposition principle holds true, thus linear combinations of these expressions are also valid solutions to the problem investigated. For the physical significance of the problem, we also require the water surface fluctuation to be real.

{\itshape{Example 3: Parabolic Bathymetry, $h_{par}(x)=c_1 x^a=c_1 x^2$, and Linearly Varying Breadth, $b_{lin}(x)=c_2 x^c=0.05 x^1$ for various values of $c_1$:}}

\begin{figure}[htb!]
\begin{center}
   \includegraphics[width=6.9in]{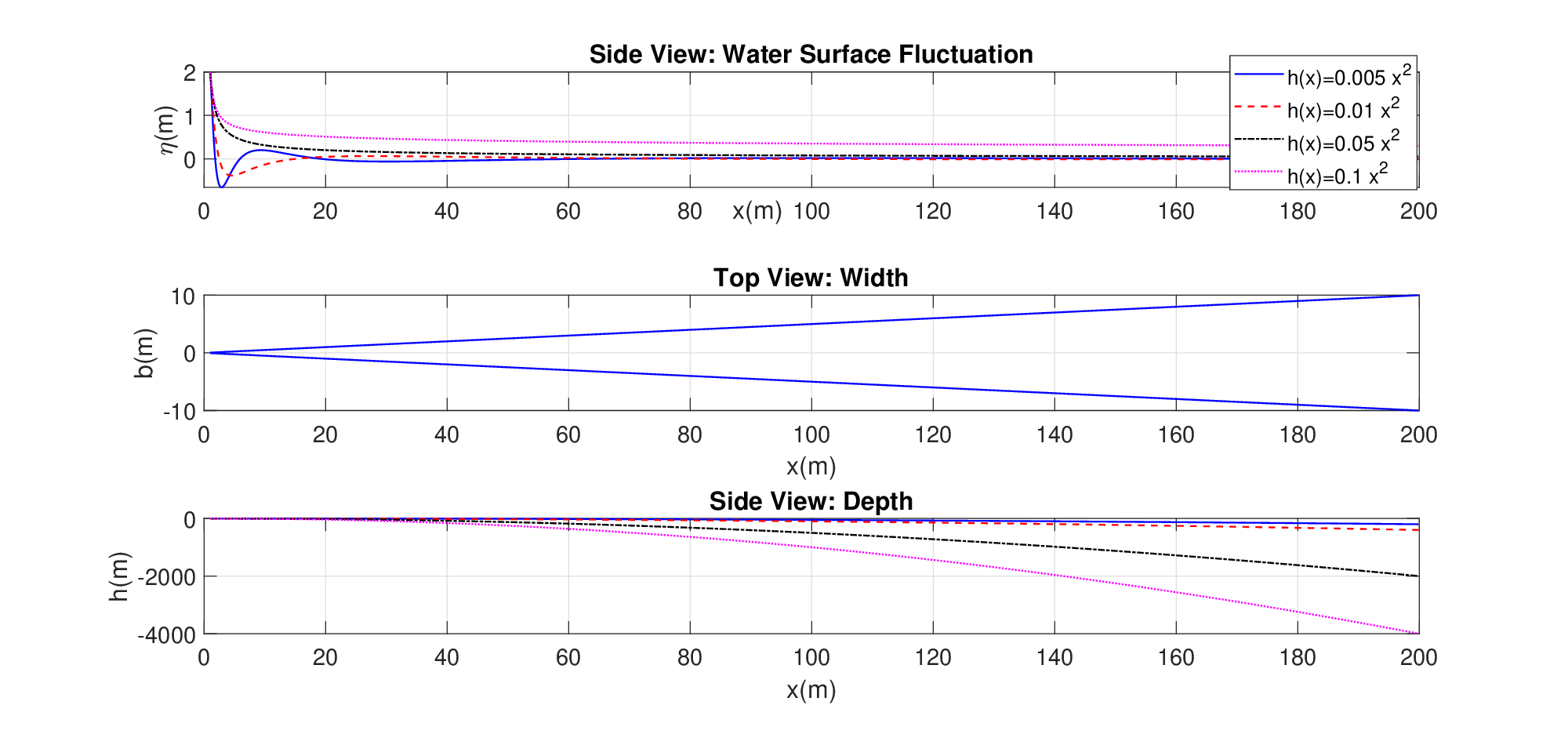}
  \end{center}
\caption{\small Long-wave solution for 4 different parabolic depth profiles and uniformly varying breadth profile for a wave period of $T=10s$ at time $t=0$. a) Water surface fluctuation b) Breadth variation c) Depth variation.}
  \label{fig3}
\end{figure}

In Fig.~\ref{fig3}, we depict the solution of the long-wave problem for 4 different parabolic depth profiles and linearly varying breadth. The results are plotted for a wave period of $T=10s$. 

As indicated in Fig.~\ref{fig3}, we observe more dips and local crests with larger oscillation amplitudes for shallower parabolic bathymetries. There is also a slight shift in the dips observed in the water surface fluctuation in the nearshore region. For deeper parabolic bathymetries, the amplitudes of oscillation on $x$ are very insignificant, whole wave field exhibits a Helmholtz mode characteristic.

{\itshape{Example 4: Parabolic Bathymetry, $h_{par}(x)=c_1 x^a=0.005 x^2$, and Various Linear and Nonlinear Varying Breadth Profiles in the form of $b(x)=c_2 x^c=0.01 x, 0.1 x, 1x,0.01 x^2, 0.01 x^3$:}}

\begin{figure}[htb!]
\begin{center}
   \includegraphics[width=6.9in]{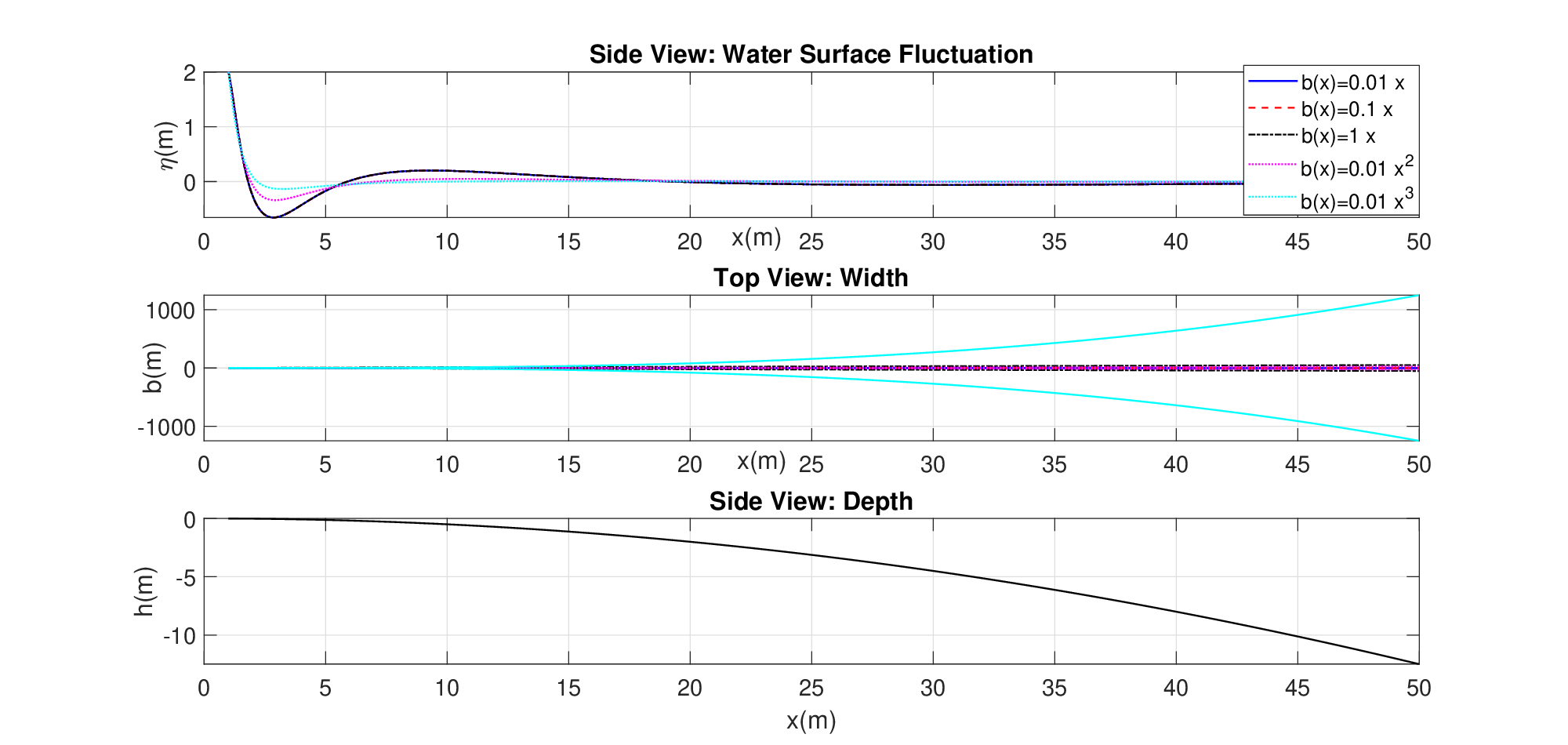}
  \end{center}
\caption{\small Long-wave solution for a parabolic depth profile with 5 different parabolic breadth profiles for a wave period of $T=10s$ at time $t=0$. a) Water surface fluctuation b) Breadth variation c) Depth variation.}
  \label{fig4}
\end{figure}

Lastly we present the solution of the long-wave equation over a parabolic depth profile of $h_{par}(x)=c_1 x^a=0.005 x^2$ with 5 different linear and nonlinear breadth profiles described by the functions $b(x)=c_2 x^c=0.01 x, 0.1 x, 1x,0.01 x^2, 0.01 x^3$ which are plotted in Fig.~\ref{fig4}. 

Fig.~\ref{fig4} confirms that as the breadth of the estuary becomes larger, the oscillation amplitudes decrease and the wavelength increases similar to the case of Bessel function solutions given and discussed in Fig.~\ref{fig2}. However, no shifts in the locations of the peaks or dips in the water surface fluctuation are observed.

%%%%%%%%%%%%%%%%%%%%%%%%%%%%%%%%%%%%%%%%%%%%%%%%%%%%%%%%%%%%%%%%

\subsection{\label{sec4:level3}Resonance of the co-oscillating tides in a channel with nonlinear depth and breadth variations in the power-law form}

Considering that the water surface elevation found in Eq.~\ref{eq03} should give a finite solution at $x=0$ for physical significance and requiring that $\frac{\sigma^2}{c_1g}>0$, Bessel-Z equation can be reduced to Bessel function of the first kind of order $|\nu|$, $J_{|\nu|}$. For an unit amplitude, the time-dependent form of the equation can be written as
\begin{equation}
   \widehat{\eta}(x,t)=x^{\nu/\alpha} J_{|\nu|}\left(\alpha \sqrt{\left|\frac{\sigma^2}{c_1g}\right|} x^{1/\alpha}\right) \cos(\sigma t),
   \label{eq07}
\end{equation}
where $\alpha=\frac{2}{2-a}$ and $\nu=\frac{1-a-c}{2-a}$. Matching the the propagating wave solution at the entrance of the channel, $x=l$, which is given by
\begin{equation}
    \widehat{\eta}(l,t)=\frac{H}{2} \cos(\sigma t),
    \label{eq08}
\end{equation}
yields
\begin{equation}
    \widehat{\eta}(x,t)=\frac{H}{2} \frac{x^{\nu/\alpha} J_{|\nu|}\left(\alpha \sqrt{\left|\frac{\sigma^2}{c_1g}\right|} x^{1/\alpha}\right)}{l^{\nu/\alpha} J_{|\nu|}\left(\alpha \sqrt{\left|\frac{\sigma^2}{c_1g}\right|} l^{1/\alpha}\right)} \cos(\sigma t),
    \label{eq9}
\end{equation}
where $H$ is the amplitude of the wave coming from offshore to the entrance of the channel. Accordingly, resonance conditions and characteristics of the problem can be analyzed by evaluating the ratio $R(x)=\left| \frac{\eta(x)}{\eta(l)} \right|$ which eventually leads to
\begin{equation}
    R(x)=\left| \frac{x^{\nu/\alpha} J_{|\nu|}\left(\alpha \sqrt{\left|\frac{\sigma^2}{c_1g}\right|} x^{1/\alpha}\right)}{l^{\nu/\alpha} J_{|\nu|}\left(\alpha \sqrt{\left|\frac{\sigma^2}{c_1g}\right|} l^{1/\alpha}\right)} \right|.
    \label{eq10}
\end{equation}
Resonant modes occur where the denominator of this function is equal to zero. Namely, the zeros of $J_{|\nu|}\left(\alpha \sqrt{\left|\frac{\sigma^2}{c_1g}\right|} l^{1/\alpha}\right)$ should be found to determine resonant frequencies of different modes.

{\itshape{Example 5: Comparison between Uniform Depth, $h_{uni}(x)=h_0$, and Linearly Varying Breadth, $b_{lin}(x)=c_2x$, Case with Equilibrium Beach Profile, $h_{eq}(x)=c_1 x^{a}=A x^{2/3}$, and Linearly Varying Breadth, $b_{lin}(x)=c_2x$, Case:}}\\
For $a=0$ and $c=1$, the problem reduces to the well-known long-wave solution for the constant depth and pie shaped beach \cite{Lamb, DeanDal} and parameters of the Bessel-Z function become $\alpha=1$ and $\nu=0$. For this scenario, the ratio $R(x)$ turns out to be in term of the Bessel functions of the first kind of order zero, which can be formulated as \cite{Lamb, DeanDal}
\begin{equation}
    R(x)=\left| \frac{J_0\left(\sqrt{\left|\frac{\sigma^2}{c_1g}\right|}x\right)}{J_0\left(\sqrt{\left|\frac{\sigma^2}{c_1g}\right|}l\right)} \right |.
    \label{eq11}
\end{equation}
For the second case of equilibrium beach profile with linearly varying breadth, parameters become $a=\frac{2}{3}$, $c=1$, and accordingly, $\alpha=\frac{3}{2}$ and $\nu=-\frac{1}{2}$. The resonance condition for this case can be formulated by setting $\alpha \sqrt{\left|\frac{\sigma^2}{c_1g}\right|} l^{1/\alpha}=z_n$, where $z_n$ are the zeroes of the Bessel function, $J_{|\nu|}$. The values that make $J_0$ and $J_{1/2}$ zero are obtained from Abramowitz and Stegun \cite{Abramowitz}. In order to illustrate the typical resonance conditions in an example, $c_1$ values are selected as $3.4200$, $0.1000$ and $0.0171$ for the cases of constant depth, equilibrium beach profile (note that $c_1=A$ for equilibrium beach profile case) and uniformly varying depth, respectively, so that maximum depth for each of these profiles is $h_{1}=3.42m$. The domain of the problem for this example is selected to be $x \in [0m, 200m]$, thus the value of $l=200m$ is used as a representative channel length. Accordingly, angular frequencies of first 5 resonant modes for this three cases are numerically calculated and tabulated in Tab.~\ref{tab1}.

\begin{table}[h]
   \centering
   \caption{Resonant modes and angular frequencies ($rad.s^{-1}$).}
   \label{tab1}
\begin{tabular}{ccclll}
\hline
Resonant modes                                                                           & 1 & \ 2 & \ \ \ 3 &  \ \ \ 4 & \ \ \ 5 \\ \hline
\begin{tabular}[c]{@{}c@{}}Resonance frequency ($rad.s^{-1}$) \\  for $h_{uni}(x)=h_0$ and $b_{lin}(x)=c_2x$\end{tabular} & 0.0696 & 0.1599 & 0.2506 & 0.3415 & 0.4324 \\
\multicolumn{1}{l}{}                                                                          & \multicolumn{1}{l}{} & \multicolumn{1}{l}{} &        &        &        \\
\begin{tabular}[c]{@{}c@{}}Resonance frequency ($rad.s^{-1}$) \\  for $h_{eq}(x)=Ax^{2/3}$ and $b_{lin}(x)=c_2x$\end{tabular} & 0.0607 & 0.1213 & 0.1820 & 0.2426 & 0.3033 \\
\multicolumn{1}{l}{}                                                                          & \multicolumn{1}{l}{} & \multicolumn{1}{l}{} &        &        &        \\
\begin{tabular}[c]{@{}c@{}}Resonance frequency ($rad.s^{-1}$) \\  for $h_{eq}(x)=c_1x$ and $b_{lin}(x)=c_2x$\end{tabular} & 0.0555 & 0.1016 & 0.1473 & 0.1929 & 0.2385
\end{tabular}
\end{table}

\begin{figure}[htb!]
\begin{center}
   \includegraphics[width=6.9in]{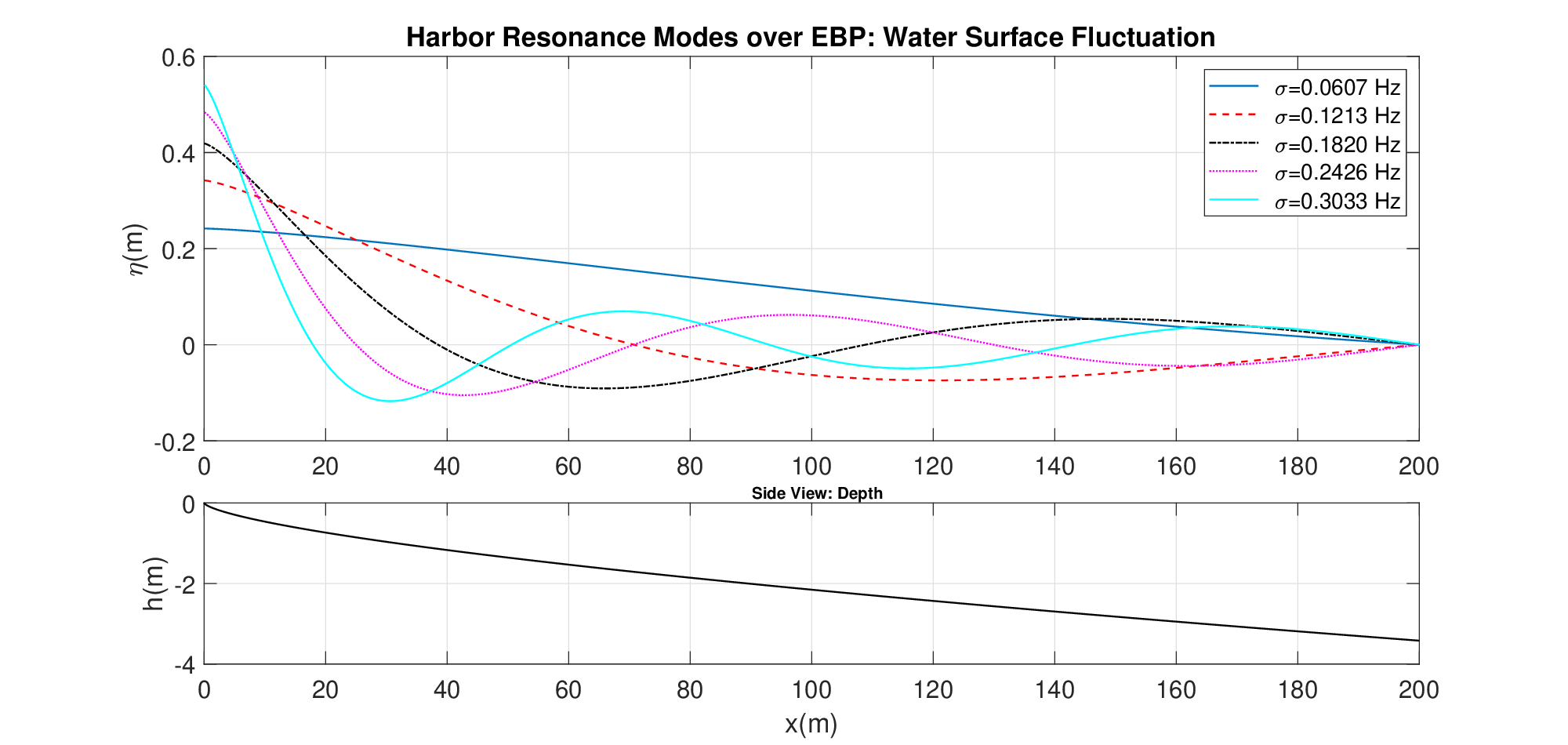}
  \end{center}
\caption{\small First five modes of harbor resonance over an equilibrium beach profile with resonant frequencies given in the second row of Tab.~\ref{tab1} at time $t=0$. a) Water surface fluctuation b) Depth variation.}
  \label{Newfig_afterrevision1}
\end{figure}

As can be seen in the Tab.~\ref{tab1}, resonant frequencies for equilibrium beach profile are smaller than the ones for the constant depth, for the same breadth shape of $b(x)=c_2x$. This is an expected result since in larger average depths, the shallow water celerity $V=\sqrt{gh}$ is larger, thus the angular frequency, $\sigma=\frac{2 \pi}{T}=\frac{2 \pi V}{L}$ is larger. It is also possible to compute the resonant wavelengths by solving the dispersion relation. The water surface profiles for the first five modes of harbor resonance over the equilibrium beach profile with angular frequencies given in the second row of Tab.~\ref{tab1} are calculated by Eq.~\ref{eq03} and depicted in Fig.~\ref{Newfig_afterrevision1}.

\subsection{\label{sec:level3} Seiching period for nonlinear depth variation in the power-law form}
\begin{figure}[htb!]
\begin{center}
   \includegraphics[width=4.0in]{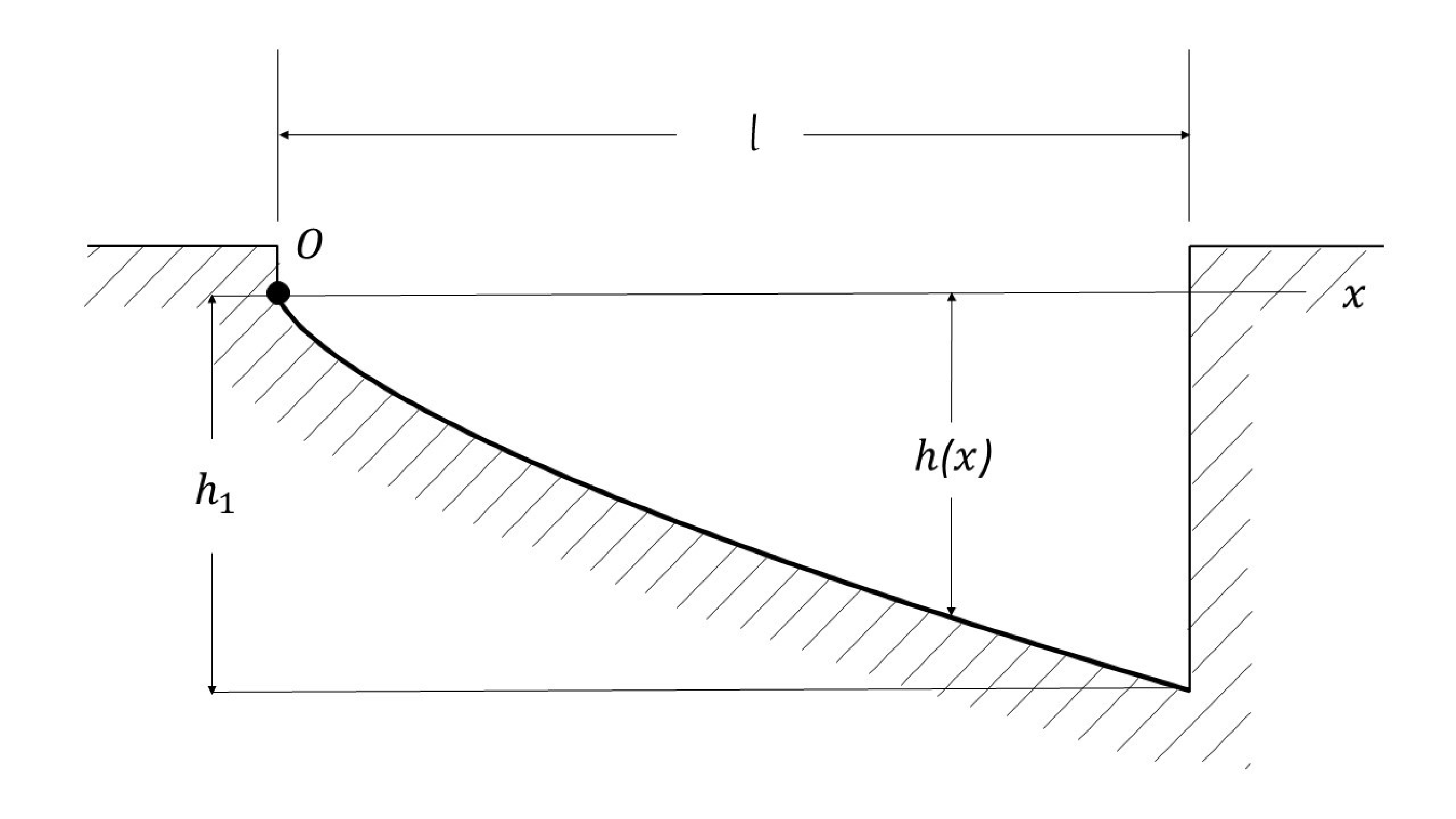}
  \end{center}
\caption{\small Equilibrium beach profile with $h_{eq}=Ax^{2/3}$.}
  \label{fig5}
\end{figure}

Transient solution of the long-wave equations given by Eq.~\ref{eq03} can be obtained using simple time harmonics which leads to
\begin{equation}
    \widehat{\eta}(x,t)=x^{\nu/\alpha} J_{|\nu|}\left(\alpha \sqrt{\frac{\sigma^2}{c_1g}}x^{1/\alpha}\right) \cos(\sigma t),
    \label{eq12}
\end{equation}
where the form is derived for constant breadth, $b$, for the case of $\frac{\sigma^2}{c_1g}>0$. For constant breadth, the power parameter of the breadth variation is $c=0$, thus the parameters read as $\alpha=\frac{2}{2-a}$ and $\nu=\frac{1-a}{2-a}$. The seiching period for this wave can be obtained by checking the minimum of the $\eta$ function. For this reason, $\frac{\partial \eta}{\partial x}$ should be set equal to zero. Inserting $\alpha=\frac{2}{2-a}$ and $\nu=\frac{1-a}{2-a}$, and evaluating the derivative one can obtain
\begin{equation}
    \frac{\partial \eta}{\partial x}= \sqrt{\frac{\sigma^2}{c_1g}} x^{\frac{1-2a}{2}} J_{\left| \frac{1-a}{2-a} \right|-1} \left(\frac{2}{2-a} \sqrt{\frac{\sigma^2}{c_1g}}x^{\frac{2-a}{2}} \right).
    \label{eq15}
\end{equation}
The fundamental seiching periods for different $a$ values are calculated by evaluating the zeroes of the function at $x=l$ given by Eq.~\ref{eq15} numerically in terms of the wavelength $L=2l$, $g$ and $c_1$. The corresponding seiching period becomes $T=\mu\frac{2l}{\sqrt{g h_1}}$. Different $\mu$ values corresponding to different $a$ values are tabulated in Tab.~\ref{tab2} for the case where the parameters illustrated in Fig.~\ref{fig5} is taken as $h_1=20m$ and $l=200m$.

\begin{table}[h]
   \centering
   \caption{Coefficients of fundamental seiching periods for different values of $a$.}
   \label{tab2}
   \begin{tabular}{ccc}
     \hline
     a  & $c_1$ & $\mu$ \\
     \hline
     0 & 20.0000 & 1.0000 \\
     0.10 & 11.7741 & 1.0403 \\
		 0.20 & 6.9314 & 1.0849 \\
		 0.30 & 4.0806 & 1.1316 \\
		 0.40 & 2.4022 & 1.1837 \\
     0.50 & 1.4142 & 1.2409 \\
		 0.60 & 0.8326 & 1.3040 \\		
     2/3 & 0.5848 & 1.3499 \\
		 0.70 & 0.4901 & 1.3740 \\
		 0.80 & 0.2885 & 1.4522 \\
		 0.90 & 0.1699 & 1.5402 \\
     1.00 & 0.1000 & 1.6398 
\end{tabular}
\end{table}
As it can be seen in Tab.~\ref{tab2}, the fundamental seiching period for the equilibrium beach profile depicted in Fig.~\ref{fig5} defined by the function $h_{eq}=Ax^{2/3}$ is found out to be $T \approx 1.3499\frac{2l}{\sqrt{g h_1}}$ for the case of $A=0.1m^{1/3}$. 

The coefficient $\mu$ is independent of $c_1$, however, $T$ depends on $c_1$ since it control the depth $h_1$. As expected, when the value of $a=1$ is used, depth variation becomes linear thus the well-known value of $T \approx 1.6398\frac{2l}{\sqrt{g h_1}}$ is obtained. Besides, at $a=0$, again the expected value of $T=\frac{2l}{\sqrt{g h_1}}$ is found. Additionally, as the parameter $a$ increases, the average depth decreases which leads to higher fundamental seiching periods. This result is also commensurate with previous findings of the literature. It is also possible to find the seiching periods for higher modes by evaluating the other zeros of the expression given by Eq.~\ref{eq15}.  
%%%%%%%%%%%%%%%%%%%%%%%%%%%%%%%%%%%%%%%%%%%%%%%%%%%%%%%%%%%%%%%%
%%%%%%%%%%%%%%%%%%%%%%%%%%%%%%%%%%%%%%%%%%%%%%%%%%%%%%%%%%%%%%%%

\section{\label{sec:level1}Conclusion and Future Work}
In this paper, we have derived the exact analytical solutions of the long-wave equation for coastal regions having a nonlinear depth and breadth variation in the power-law form. We have shown that the solutions of the long-wave equation in such regions can be described by Bessel-Z functions. We depicted and discussed the applicability and possible usage of our findings, including the analysis of wavefields over equilibrium beach profiles. Additionally, we have developed the seiching periods and possible resonance conditions over bathymetries where such power-law variation can be observed. In addition to the analysis of the wavefields over the nonlinear depth and breadth described above, our results can also be used to investigate harbor resonance and shift in resonant periods due to reconstruction, sedimentation, and dredging. Hydrodynamics in the vicinity of sufficiently long seawalls with linear or power-law nonlinear layout variations together with bathymetric variations and around energy focusing structures and islands, wave reflection at inward corners, runup on revetments such as those discussed in \cite{Goda}, can also be investigated in the frame of our analytical findings. Kirby et al. \cite{KirbyHaller} showed that low frequency motions which are indicative of wave seiching are also observed in laboratory flume experiments, thus our results can be used to investigate long-wave and seiching phenomena in laboratory environment for various beach profiles including equilibrium beach profiles.

In a more general setting, our results can be used to estimate the resonance frequencies of the acoustic and optical resonators, when different geometries such as prismoidal shapes are used to resonate sound or light waves \cite{Auld, Lakinacousticresonator, LigeometryResonator, Chevalieropticalresonator}. The effects of doping, accretion and erosion of materials such as dust (see for example \cite{FerreiraOliveira, KimChoi, Matsko}) can also be investigated within the analytical frame we presented. 

In near future, we aim to expand our analysis to various problems of great importance for the ocean modeling community.
We plan to extend our findings to the waves with frictional damping and under the geostrophic effects. We also aim to investigate the solutions of the mild-slope equation proposed by Berkhoff \cite{Berkhoff} which can be formulated as	
\begin{equation}
\nabla \cdot (p \nabla \eta)+\sigma^2 q \eta=0,
\label{eq1_sec2}
\end{equation}
where $\sigma$ is the angular wave frequency, $h(x,y)$ is the depth, $q=\frac{1}{2} \left(1+ \frac{2kh}{\sinh 2kh}\right)=\frac{C_g}{C}$ and $p=gh \frac{\tanh kh}{kh}q=CC_g$ where $C$ denotes the celerity, $C_g$ denotes the group velocity, with the dispersion relationship given by $\sigma^2=gk\tanh kh$ to the cases where depth and breadth variations obey nonlinear power-law forms. Additionally, the 2D version of the long-wave equation given by Eq.~(\ref{eq01}) is treated in \cite{Kirby1981, Ursell} for analysis of the edge waves. It remains an open question that how the edge waves behave over nonlinear depth and breadth variations in the power-law form. Possible Bessel-Z analogs of the Laguerre polynomials and their reducibility will be investigated for this purpose. It is also possible to extend our analysis for the long-waves presented for various nonlinear depth and breadth profiles in the power-law form to obtain solutions in terms of Laguerre, Chebyshev, Hermite, or Legendre polynomials \cite{Arfken, Whittaker, HilbertCourant} and their possible reducible extensions to their self-adjoint forms. Such extensions would be useful for analyzing the effects of various harbor inlet shapes on harbor circulations and hydrodynamics, as well as for analysis of sound and light resonators having various geometries. Another possible research direction to follow is to extend the tsunami runup calculations such as those presented in \cite{Synolakis} and extend the similar solitary waves studies \cite{Inc1, Inc2, BayRINP, Inc3, Inc4, BayCNSNS, Inc5, Bay2021DOA} and rogue wave studies \cite{Kharif, BayPLA, BayPRE1, BayPRE2} to take the effects of equilibrium beach profiles into consideration with solutions given in terms of Bessel-Z functions and analyze the effects of realistic profiles on tsunami runup and inundation. Investigation of fully nonlinear wave characteristics and their evolution over equilibrium beach profiles using numerical approaches such as those discussed in \cite{Baysci, Karjadi2012} is also planned.

\end{document}